\documentclass[twocolumn]{article}

\usepackage{amsmath}
\usepackage{amssymb}
\usepackage{amsthm}
\usepackage{graphicx}
\usepackage{framed}
\usepackage{url}
\usepackage{listings}
\usepackage{algorithm}
\usepackage{algorithmic}
\usepackage{times}
\usepackage{fullpage}

\newtheorem{definition}{Definition}
\newtheorem{lemma}{Lemma}

\newtheorem{conjecture}{Conjecture}
\newtheorem{corollary}{Corollary}
\newtheorem{observation}{Observation}

\newcommand{\figwidth}{0.85\columnwidth}
\newcommand{\comment}[1]{}

\title{Mining Statistically Significant Substrings Based on the Chi-Square Measure}

\author{
\hspace*{-5mm}
\begin{tabular}{cc}
	Sourav Dutta & Arnab Bhattacharya \\
	\url{sdutta@iitk.ac.in} & \url{arnabb@iitk.ac.in} \\
	\multicolumn{2}{c}{Dept. of Computer Science and Engineering,} \\
	\multicolumn{2}{c}{Indian Institute of Technology, Kanpur} \\
	\multicolumn{2}{c}{Kanpur, UP 208016, India.} \\
\end{tabular}
}

\date{}

\begin{document}

\maketitle
\begin{abstract}
	Given the vast reservoirs of data stored worldwide, efficient mining of data 
	from a large information store has emerged as a	great challenge.  Many databases 
	like that of intrusion detection systems, web-click records, player statistics, 
	texts, proteins etc., store strings or sequences.  Searching for an 
	unusual pattern within such long strings of data has emerged as a requirement 
	for diverse applications.  Given a string, the problem then is to identify the 
	substrings that differs the most from the expected or normal behavior, i.e., 
	the substrings that are statistically significant.  In other words, these 
	substrings are less likely to occur due to chance alone and may point to some 
	interesting information or phenomenon that warrants further exploration.  To 
	this end, we use the chi-square measure.  We propose two heuristics for retrieving 
	the top-k substrings with the largest chi-square measure.  We show that the 
	algorithms outperform other competing algorithms in the runtime, while maintaining 
	a high approximation ratio of more than 0.96.
\end{abstract}

\section{Motivation}
\label{sec:motiv}

A recent attractive area of research has been the detection of statistically
relevant sequences or mining interesting patterns from 
within a given string~\cite{bioinfo,anomaly}.  Given an input string composed of
symbols from a defined alphabet set with a probability distribution defining
the chance of occurrence of the symbols, and thereby defining its expected
composition, we would like to find the portions of the string which
deviate from the expected behavior and can thus be potent sources of study for hidden
pattern and information.  An automated monitoring system like a cluster of
sensors sensing the temperature of the surrounding environment for fire alert, or a
connection server sniffing the network for possible intrusion detection provides
a few of the applications where such pattern detection is essential.  Other applications 
involve text analysis of e-mails and blogs to predict terrorist activities or judging
prevalent public sentiments, studying trends of the stock market, and identifying
sudden changes in the mutation characteristics of protein sequence of an
organism.  Similarly, information extracted from a series of Internet websites 
visited, the advertisements clicked on them or from the nature of transactions on a
database, can capture the interests of the end user, prospective clients and also the periods of heavy
traffic in the system.  An interesting field of application can be the identification of 
good and bad career patches of a sports icon.  For example, given the runs scored by Sachin 
Tendulkar in each innings of his one-day international cricket career, we may 
be interested in finding his in-form and off-form patches. 

Quantifying a substring or an observation as unexpected under a given circumstance relies 
on the probabilistic analysis used to model the deviation of the behavior from its 
expected nature.  Such an outcome that deviates from the expected, then becomes interesting 
and may reveal certain information regarding the source and nature of the variance, 
and we are interested in detecting such pockets of hidden data within substrings of 
an input string.  A statistical model is used to determine the relationship of an 
experimental or observed outcome with the factors affecting the system, or to establish 
the occurrence as pure chance.  An observation is said to be statistically significant
if its presence cannot be attributed to randomness alone.  For example, within a large DNA 
sequence, the recognition of hugely variational patterns involve probability 
matching with large fluctuations, thereby the need to predict the locations 
uses self-consistent statistical procedures~\cite{mot}.  

The degree of uniqueness of a pattern can be captured by several measures
including the p-value and z-score~\cite{dp,crit}.  For evaluating the
significance of a substring, it has been shown that the p-value provides a more
precise conclusion as compared to that by the z-score~\cite{bioinfo}.  However,
computing the p-value entails enumerating all the possible outcomes, which can
be exponential in number, thus rendering it impractical.  So, heuristics based
on branch-and-bound techniques have been proposed~\cite{categ}.  The
\emph{log--likelihood ratio}, $G^2$~\cite{multi} provides such a measure based
on the extent of deviation of the substring from its expected nature.  For
multinomial models, the $\chi^2$ statistic approximates the importance of a
string more closely than the $G^2$ statistic~\cite{multi,pear}.  Existing
systems for intrusion detection use multivariate process control techniques such
as Hotelling's $T^2$ measure~\cite{hotel}, which is again computationally
intensive.  The chi-square measure, on the other hand, provides an easy way to
closely approximate the p-value of a sequence~\cite{multi}.  To simplify
computations, the $\chi^2$ measure, unlike Hotelling's method, does not consider
multiple variable relationship, but is as effective in identifying ``abnormal''
patterns~\cite{anomaly}.  Thus, in this paper, we use the Pearson's $\chi^2$
statistic as a measure of the p-value of a substring~\cite{multi,pear}.  The
$\chi^2$ distribution is characterized by the \emph{degrees of freedom}, which
in the case of a string, is the number of symbols in the alphabet set minus one.
The larger the $\chi^2$ value of a string, the smaller is its p-value, and hence
more is its deviation from the expected behavior.  So, essentially our problem
reduces to finding the substring that has the maximum $\chi^2$ value.  We
propose to extract such substrings efficiently.\\

\noindent
\textbf{Related Work:}\\
Formally, given a string $S$ composed of symbols from the alphabet set $\Sigma$
with a given probability distribution $P$ modeling the chance of occurrence of
each symbol, the problem is to identify and extract the top-$k$ substrings
having the maximum chi-square value or the largest deviation within the
framework of p-value measure for the given probability distribution of the
symbols.  Na\"\i vely we can compute the $\chi^2$ value of all the substrings
present in $S$ and determine the top-k substrings in $O(l^2)$ time for a string
of length $l$ (see Algorithm~\ref{alg:naive}).  The blocking algorithm and its
heap variant proposed in~\cite{sumit}, reduce the practical running time for
finding such statistically important substrings, but suffers from a high
worst-case running time.  The number of blocks found by this strategy increases
with the size of the alphabet set and also when the probabilities of the
occurrence of the symbols are nearly similar.  In such scenarios, the number of
blocks formed can be almost equal to the length of the given string, thereby
degenerating the algorithm to that of the na\"\i ve one.  The heap variant
requires a high storage space for maintaining the separate \emph{max} and
\emph{min} heap structures and also manipulates a large number of pointers.
Further, the algorithm does not easily generalize beyond static input strings,
and cannot handle top-k queries.  In time-series databases, categorizing a
pattern as surprising based on its frequency of occurrence and mining it
efficiently using suffix trees has been proposed in~\cite{keo}.  However, the
$\chi^2$ measure, as discussed earlier, seems to provides a better parameter for
judging whether a pattern is indeed interesting.   

\begin{algorithm}[t]
\caption{Na\"\i ve Algorithm}
\label{alg:naive}
\begin{algorithmic}[1]
\REQUIRE String $S$ with the probability of occurrence of each symbol in the alphabet set.
\ENSURE Top-k substrings having the maximum $\chi^2$ value.
\STATE Extract all the substrings in $S$.
\STATE Compute the $\chi^2$ value of all the substrings.
\STATE Return the substrings having the top-k $\chi^2$ value.
\end{algorithmic}
\end{algorithm}

In this paper, we propose two algorithms, \emph{All-Pair Refined Local Maxima
Search} (ARLM) and \emph{Approximate Greedy Maximum Maxima Search} (AGMM) to
efficiently search and identify interesting patterns within a string.  
We show that the running time of the algorithms are far better than the existing
algorithms with lesser space requirements.  The procedures can also be
easily extended to work in streaming environments.  ARLM, a
quadratic algorithm in the number of local maxima found in the input string,
and AGMM, a linear time algorithm, both use the presence of local maxima in the
string.  We show that the 
approximation ratio of the reported results to the actual is 0.96 or more.
Empirical results emphasize that the algorithms work efficiently.  

The outline of the paper is as follows:
Section~\ref{sec:def} formulates the properties and behavior of strings
under the $\chi^2$ measure.  Section~\ref{sec:algo} describes the two proposed
algorithms along with their runtime complexity analysis.  Section~\ref{sec:expt}
shows the experimental results performed on real and synthetic data, before 
Section~\ref{sec:conc} concludes the paper.

\section{Definition and Properties}
\label{sec:def}

Let $str = s_1s_2 \dots s_l$ be a given string of length $l$ composed of symbols $s_i$ taken from the 
alphabet set $\Sigma = \{\sigma_1,\sigma_2, \dots ,\sigma_m\}$, where $|\Sigma| = m$.  To each symbol 
$\sigma_i \in \Sigma$ is associated a $p_{\sigma_i}$ (henceforth represented as $p_i$), denoting 
the probability of occurrence 
of that symbol, such that $\sum^m_{i=1}p_i = 1$.  Let $\theta_{\sigma_i,str}$ (henceforth represented as 
$\theta_{i,str}$) denote the 
observed number of the symbol $\sigma_i$ in the string $str$, where $\sigma_i \in \Sigma$ and $str \in \Sigma^*$.

The chi-square value of a string $str \in \Sigma^*$ of length $l$ is computed as 
\begin{align}
	\label{eq:chi}
	\chi^2_{str} &= \sum^m_{i=1} \frac {\left(p_il - \theta_{i,str}\right)^2} {p_il}
\end{align}
The chi-square measure thus calculates the deviation of the composition of the string from its 
expected nature by computing the sum of the normalized square of difference of the observed value 
of each symbol in the alphabet set from the expected value of occurrence. 

\begin{observation} 
	Under string concatenation operation (.), for two arbitrary strings $a$ and $b$ 
	drawn from the same alphabet set and probability distribution of the symbols (henceforth referred to 
	as the \emph{same universe}), the $\chi^2$ measure of the concatenated string is commutative in the order 
	of concatenation.
\end{observation}

\begin{proof}
	It is easy to observe that the lengths of $a.b$ and $b.a$ are the same.  Further, the 
	observed values of the different symbols and their probabilities of occurrence are the 
	same in both the concatenated strings.  Hence, the $\chi^2_{a.b}$ is equal to $\chi^2_{b.a}$ 
	according to Eq.~\eqref{eq:chi}. 
\end{proof}

\begin{lemma}
	The $\chi^2$ value of the concatenation of two strings drawn from the same universe is less than 
	or equal to the sum of the $\chi^2$ values of the individual strings.
\end{lemma}

\begin{proof}
	Let $a$ and $b$ be two strings, of length $l_a$ and $l_b$ respectively.  Let $a.b$ form the 
	concatenated string having length $(l_a+l_b)$.
	Using Eq.~\eqref{eq:chi}, the sum of the chi-square values of the strings is
	\begin{align}
		\label{eq:a&b}
		&\chi^2_a + \chi^2_b = \sum^m_{i=1}\left( \frac{\left(p_il_a - \theta_{i,a}\right)^2}{p_il_a} + 
		\frac{\left(p_il_b - \theta_{i,b}\right)^2}{p_il_b}\right) \\
		&\text{Now, }
		\label{eq:ab}
		\chi^2_{ab} = \sum^m_{i=1}\frac{\left(p_i\left(l_a+l_b\right)-\theta_{i,{ab}}\right)^2}{p_i
		\left(l_a+l_b\right)} \\
		&\text{Using $\theta_{i,ab} = \theta_{i,a} + \theta_{i,b}$ and Eqs.~\eqref{eq:a&b} and~\eqref{eq:ab}, 
		we have } \nonumber \\
		&\chi^2_a + \chi^2_b - \chi^2_{ab} = \sum^m_{i=1}\left( \frac{\left(p_il_a - \theta_{i,a}\right)^2}
		{p_il_a} + \right. \nonumber \\
		&\left. \qquad \qquad \frac{\left(p_il_b - \theta_{i,b}\right)^2}{p_il_b} - 
		\frac{\left(p_i\left(l_a+l_b\right)-\theta_{i,{ab}}\right)^2}{p_i\left(l_a+l_b\right)}\right) \nonumber \\
		&= \sum^m_{i=1}\left( \frac{\left(p_il_a - \theta_{i,a}\right)^2}{p_il_a} + 
		\frac{\left(p_il_b - \theta_{i,b}\right)^2}{p_il_b} - \right. \nonumber \\
		&\left. \qquad \qquad \qquad \quad \frac{\left(\left(p_il_a-\theta_{i,a}\right) + 
		\left(p_il_b - \theta_{i,b}\right)\right)^2}{p_i\left(l_a+l_b\right)}\right) \nonumber
	\end{align}
	\begin{align}
		&= \sum^m_{i=1}\left( \frac{\left(p_il_a - \theta_{i,a}\right)^2}{p_i} \left[\frac{1}{l_a} 
		- \frac{1}{l_a+l_b}\right] + \right. \nonumber \\
		&\left. \qquad \qquad \quad \frac{\left(p_il_b-\theta_{i,b}\right)^2}{p_i }\left[\frac{1}{l_b} - 
		\frac{1}{l_a+l_b}\right] - \right. \nonumber \\
		&\left. \qquad \qquad \qquad \qquad \quad 2\frac{\left(p_il_a - \theta_{i,a}\right)\left(p_il_b - 
		\theta_{i,b}\right)}{p_i\left(l_a + l_b\right)} \right) \nonumber \\
		&= \sum^m_{i=1}\left( \frac{\left(p_il_a - \theta_{i,a}\right)^2l_b}{p_il_a\left(l_a + 
		l_b\right)} + \frac{\left(p_il_b - \theta_{i,b}\right)^2l_a}{p_il_b\left(l_a + l_b\right)} - 
		\right. \nonumber \\
		&\left. \qquad \qquad \qquad \qquad \quad 2\frac{\left(p_il_a - \theta_{i,a}\right)\left(p_il_b - 
		\theta_{i,b}\right)}{p_i\left(l_a + l_b\right)} \right) \nonumber \\
		\quad &= \frac{1}{l_al_b} \sum^m_{i=1}\left( \frac{\left(p_il_a - 
		\theta_{i,a}\right)^2l^2_b}{p_i\left(l_a + l_b\right)} + \frac{\left(p_il_b - 
		\theta_{i,b}\right)^2l^2_a}{p_i\left(l_a + l_b\right)} - \right. \nonumber \\
		&\left. \qquad \qquad \qquad \quad 2\frac{l_al_b\left(p_il_a - 
		\theta_{i,a}\right)\left(p_il_b -\theta_{i,b}\right)}{p_i\left(l_a + l_b\right)} \right) 
		\nonumber \\
		&= \frac{1}{l_al_b} \sum^m_{i=1}\left(\frac{\left(p_il_a - \theta_{i,a}\right)l_b}{\sqrt{p_i
		\left(l_a + l_b\right)}} - \frac{\left(p_il_b - \theta_{i,b}\right)l_a}{\sqrt{p_i\left(l_a + 
		l_b\right)}} \right)^2 \nonumber \\
		&\geq 0 \nonumber
	\end{align}
	Therefore, $\chi^2_a + \chi^2_b \geq \chi^2_{ab}$.
\end{proof}

\begin{lemma}
	\label{lem:single}
	The chi-square value of a string composed of only a single type of symbol increases 
	with the length of the string.
\end{lemma}

\begin{proof}
	Let $str$ be a string of length $l$ composed only of the symbol $\sigma_j$, drawn from 
	the alphabet set $\Sigma$.  Here, $\theta_{i,str} = 0, \forall i \in \{1,2, \dots ,m\}, 
	i\neq j$ and $\theta_{j,str} = l$, as $str$ consists only $\sigma_j$. 
	Substituting the values in Eq.~\eqref{eq:chi}, we have
	\begin{align}
		\label{eq:len_l}
		\chi^2_{str} &= \frac{\left(p_j-1\right)^2l}{p_j} + \sum^m_{i=1,i\neq j}p_il 
	\end{align}
If the length of $str$ is increased by one, by including another $\sigma_j$, its chi-square value becomes

\begin{align}
	&\chi^2_{str^\prime} = \frac{\left(p_j-1\right)^2\left(l+1\right)}{p_j} + 
	\sum^m_{i=1,i\neq j}p_i\left(l+1\right)  \nonumber \\
	\label{eq:len_l1}
		&= \frac{\left(p_j-1\right)^2l}{p_j} + \frac{\left(p_j-1\right)^2}{p_j} + 
	\sum^m_{i=1,i\neq j}p_il + \sum^m_{i=1,i\neq j}p_i
\end{align}

Comparing Eq.~\eqref{eq:len_l1} with Eq.~\eqref{eq:len_l}, we observe that the chi-square value increases, since 
$p_i \geq 0, \forall i \in \{1, 2, \dots ,m\}$.
\end{proof}
With this setting, we now define the term \emph{local maxima} and describe the procedure for 
finding such a local maxima within a given string.

\begin{definition}[Local maxima]
	The \emph{local maxima} is a substring, such that while traversing through it, the inclusion 
	of the next symbol does not decrease the $\chi^2$ value of the resultant sequence.
\end{definition}

Let $s_1s_2 \ldots s_n$ be a local maxima of length $n$, where $s_i \in \Sigma, \forall i$.  
Then the following holds
\begin{align}
	\chi^2_{s_1s_2} &\geq \chi^2_{s_1} \text{ , } \chi^2_{s_1s_2s_3} \geq \chi^2_{s_1s_2} \text{ , } \dots \nonumber \\
	\chi^2_{s_1s_2 \ldots s_{n}} &\geq \chi^2_{s_1s_2 \ldots s_{n-1}} \text{ and, } 
	\chi^2_{s_1s_2 \ldots s_{n+1}} \leq \chi^2_{s_1s_2 \ldots s_n} \nonumber
\end{align}

The process of finding the local maxima involves a single scan of the entire
string.  We consider the first local maxima to start at the beginning of the given 
string.  We keep appending the next symbol to the current substring until there
is a decrease in the chi-square value of the new substring.  
The present substring is then considered to be a
local maxima ending at the previous position and the last symbol appended signifies 
the start of the next local maxima.  Thus by traversing through the entire string once, 
we find all the local maxima present, which takes $O(l)$ time for a string of length $l$.

As an example, consider the string $str=$ \emph{aaaabbba}, having $\Sigma = \{a,b\}$ with the probability of
occurrence of symbol \emph{a} as $0.2$, and that of \emph{b} as $0.8$.  
Starting from the beginning, we compute the $\chi^2$ value of \emph{a} to be
$4$.  This is considered to be the starting of a local maxima.  Appending the
next symbol, the chi-square value of \emph{aa} increases to $8$. Since the score increases, the
current local maxima is updated to \emph{aa}.  We keep appending the next symbol into
the current maxima.  We find that $\chi^2_{aaaa} = 16$
and $\chi^2_{aaaab} = 11.25$.  As there is a decrease in the chi-square value of
the substring after insertion of \emph{b}, the current local maxima becomes \emph{aaaa} 
and the next local maxima is said to begin at \emph{b}.  Repeating this procedure
for the entire string $str$, the local maxima found are \emph{aaaa}, \emph{bbb} 
and \emph{a}.

\begin{lemma}
	The expected number of local maxima present in a string of length $l$ is $O(l)$.
\end{lemma}

\begin{proof}
	From Lemma~\ref{lem:single}, we can observe that, in a local maxima if the two adjacent 
	symbols are the same, then the chi-square value cannot decrease.  Thus the current local maxima may 
	end only when a pair of adjacent symbols are different.  We would like to find the expected 
	number of positions in the string where such a boundary may exist.  Let us define an indicator variable 
	$x_i$, where $x_i=1$ if the $i^{th}$ and the $(i+1)^{th}$ symbols in the string are dissimilar, and 
	$x_i=0$ otherwise.  Let $X=\sum^{l-1}_{i=1}x_i$, where $E[X]$ gives the expected number of local maxima 
	boundaries for a string of length $l$.  $P(x_i=1)$ denotes the probability of the event $x_i=1$.  Therefore, 
	\begin{align}
		&P(x_i=1) = \sum_{\forall j,k,~j\neq k}p_jp_k = 2\times \sum_{\forall j,k,~j<k}p_jp_k \nonumber \\
		&\qquad \qquad \qquad \qquad \quad \qquad \text{[where $j,k \in \{1,2,\dots,m\}$]} \nonumber \\
		&E[X] = E[\sum^{l-1}_{i=1}x_i]= \sum^{l-1}_{i=1}E[x_i] ~~\text{[Linearity of expectation]} \nonumber \\
		&= \sum^{l-1}_{i=1}P(x_i=1) = 2 \times \sum^{l-1}_{i=1} \sum_{\forall j,k,~j<k}p_jp_k \nonumber \\
		\label{eq:expected}
		&= 2\times (l-1) \times \sum_{\forall j,k,~j<k}p_jp_k
	\end{align}
	Hence, the expected number of local maxima is $O(l)$ for a string of length $l$.
\end{proof}
However, practically 
the number of local maxima will be much less than $l$, as all adjacent positions of dissimilar symbols 
may not correspond to a local maxima boundary.  Using 
Eq.~\eqref{eq:expected}, for $m=2$ the maximum number of expected local maxima is $(l-1)/2$ and is 
$2(l-1)/3$ for $m=3$, which is obtained by substituting the maximum possible value of $P(x_i=1)$. \\

We further optimize the local maxima finding procedure by initially \emph{blocking} the string 
$str$, as described in~\cite{sumit}, and then searching for the local maxima.  This makes the 
procedure faster and concise.  A contiguous sequence of the same symbol is considered
to be a block, and is replaced by a single instance of that symbol representing
the block.  If a symbol is selected, the entire block associated with it is
considered to be selected.  

The next lemma states that if the inclusion of the symbol representing a block increases the $\chi^2$ 
value, then the inclusion of the entire block will further increase the $\chi^2$ value.  This has been 
proved in Lemma 3.2.5 and Corollary 3.2.6 on page 35-37 of~\cite{sumit}.  For completeness, we 
include a sketch of the proof in this paper.

\begin{lemma}
	If the insertion of a symbol of a block increases the chi-squared value of the 
	current substring, then the chi-squared value will be maximized if the entire block is inserted.
\end{lemma}

\begin{proof}
	Let the current substring be $sub$ and the adjacent block of length $n$ be composed of 
	symbol $\sigma_e \in \Sigma$.  Appending one $\sigma_e$ to $sub$ increases the $\chi^2$ 
	value of the new substring.  
	\begin{align}
		&\text{Given, }\sum^m_{i=1, i \neq e} \frac{\left(p_i \left(l_{sub}+1 \right) -
		\theta_{i,sub+1} \right)^2}{p_i\left(l_{sub}+1 \right)} + \nonumber \\
		&\qquad \qquad \qquad \qquad \frac{\left( p_e \left (l_{sub}+1\right) - \theta_{e,sub+1}\right)^2}{p_e \left(
		l_{sub}+1 \right)} \geq \nonumber \\
		&\qquad \qquad \sum^m_{i=1, i \neq e} \frac{\left(p_il_{sub} - \theta_{i,sub} \right)^2}{p_il_{sub}} + 
		\frac{\left( p_el_{sub} - \theta_{e,sub} \right)^2}{p_el_{sub}} \nonumber \\
		&\text{Or, } \chi^2_{sub+1} \geq \chi^2_{sub} \nonumber \\
		&\text{By simple algebraic manipulation, we can show that, } \nonumber \\
		&\chi^2_{sub+n} \geq \chi^2_{sub+n-1} \geq \dots \geq \chi^2_{sub+2} \geq \chi^2_{sub+1} \nonumber 
	\end{align}
	Hence, by including the entire block the $\chi^2$ value of the substring will be maximized.
\end{proof}
The entire string is now block-ed and the local maxima finding procedure works not 
with the original $str$ but with \emph{aba}, where the 
first \emph{a} represents the four contiguous $a$'s in $str$, the \emph{b} represents 
the next three $b$'s, and the final \emph{a} stands for the last occurrence of a.  The local 
maxima thus found are \emph{a}, \emph{b} and \emph{a}.  The positions of the 
local maxima are $1$, $5$ and $8$ respectively, according to their position in the original 
string. 

Given the position of component $y$ for each local maxima of a string, we need to extract the 
\emph{global maxima}, which we formally define as follows.

\begin{definition}[Global maxima]
\emph{Global maxima} is the substring having the maximum chi-square value, and is the 
substring that we are interested in extracting, i.e., the output substring.  
\end{definition}
The global maxima has the maximum score among all possible substrings present in the input string.  

\section{Algorithms}
\label{sec:algo}

Based on the observations, lemmas and local maxima extracting procedure discussed previously, 
in this section we explain the \emph{All-Pair Refined Local Maxima} (ARLM) and 
\emph{Approximate Greedy Maximum Maxima} (AGMM) search algorithms for mining the most 
significant substring based on the chi-square value.  
  
\subsection{All-Pair Refined Local Maxima Search Algorithm (ARLM)}
\label{ssec:allpair}

Given a string $str$ of length $l$ and composed of symbols from the alphabet set
$\Sigma$, we first extract all the local maxima present in it in linear time, as described earlier.  
We also optimize the local maxima finding procedure by
incorporating the idea of the blocking algorithm.  
With $str$ partitioned into its local maxima, the global maxima 
can either start from the beginning of a local maxima or from a position within it.  
Thus, it can contain an entire local maxima, a suffix of it or itself be a substring 
of a local maxima.  It is thus intuitive that the global maxima should begin at a 
position such that the subsequent sequence of characters offer the maximum chi-square 
value.  Otherwise, we could keep adding to or deleting symbols from the front of such 
a substring and will still be able to increase its $\chi^2$ value.  Based on this, 
the ARLM heuristic finds within each local 
maxima the suffix having the maximum chi-square value, and considers the position of 
the suffix as a potential starting point for the global maxima.

Let $xyz$ be a local maxima, where $x$ is a prefix of length $l_x$, $y$ is a
single symbol at position $pos$, and $z$ be the remaining suffix having length $l_z$.  
Categorizing the components, namely $x,y$ and $z$ of a local maxima appropriately, is 
extremely crucial for finding the global maxima.  Let $start\_pos$ and $end\_pos$ be two
lists which are initially empty and will contain the position of component $y$, 
i.e., $pos$, for each of the local maxima.  For a local maxima the chi-square value of 
all its suffices is computed.  The starting position of the suffix
having the maximum chi-square value provides the position $pos$ for the
component $y$, i.e, $yz$ will be the suffix of $xyz$ having the maximum chi-square value.  
The position $pos$ is inserted 
into the list $start\_pos$.  If no such proper suffix exists for the local maxima, 
the starting position of the local maxima $xyz$ relative to the original string 
is inserted in the list.  After populating the $start\_pos$ list with position 
entries of $y$ for each of the local maxima of the input string, the list contains 
the prospective positions from where the $global~maxima$ may start.  

The string $str$ is now reversed and the same algorithm is re-run.  This time, the
$end\_pos$ list is similarly filled with positions $y^{\prime}$ relative to the beginning 
of the string.

For simplicity and efficiency of operations, we maintain a table, $symbol\_count$
having $m$ rows and $l$ columns, where $m$ is the cardinality of the alphabet
set.  The rows of the table contain the observed number of each associated
symbols present in the length of the string denoted by the column.  The observed
count of a symbol between two given positions of the string can thus be easily
found from this table in $O(1)$ time.  The space required in this case becomes
$O(lm)$.  However, the table reduces the number of accesses of the original string 
for computing the maximum suffix within each local maxima.  It also helps to 
generalize the algorithm to streaming environments, where it is not possible to store 
the entire string.    

Given the two non-empty $start\_pos$ and $end\_pos$ lists, we now find the
chi-square value of substrings from position $g \in start\_pos$ to $h \in
end\_pos$, and $g \leq h$.  The substring having the maximum value is reported
as the global maxima.  While computing the chi-square values for all the pairs
of positions in the two list, the top-k substrings can be maintained using a
heap of $k$ elements (see Algorithm~\ref{alg:arlm} for the pseudo-code).  

\begin{algorithm}[t]
\caption{ARLM Algorithm}
\label{alg:arlm}
\begin{algorithmic}[1]
\REQUIRE String $S$ with the probability of occurrence of each symbol in the alphabet set.
\ENSURE Top-k substrings with the maximum $\chi^2$ value.
\STATE Find all the local maxima in $S$ and $S$ reversed.
\STATE $start\_pos \leftarrow$ position of suffices with maximum $\chi^2$ value in each local maxima of $S$.
\STATE $end\_pos \leftarrow$ position of suffices with maximum $\chi^2$ value in each local maxima of $S$ reversed.
\STATE Based on the $\chi^2$ value, return the top-k substrings formed from all pairs of positions from the two lists.
\end{algorithmic}
\end{algorithm}

Continuing with our example of $str=aaaabbba$, the $start\_pos$ list contains 1, 5
and 8, as the final local maxima does not contain a proper suffix with a larger 
chi-square value greater than itself.  
Computing on $str$ reversed, the $end\_pos$ list 
will contain 8, 7 and 4.  We now consider the substrings formed by the pairs
(1,8), (1,7), (1,4), (5,8), (5,7), and (8,8).  Calculating all the chi-square
values and comparing them, we find that (1,4) has the maximum value and is
reported as the global maxima which is \emph{aaaa}.  Taking $k=2$, we find that the
substring \emph{aaaabbba} corresponding to (1,8) provides the second highest
chi-square value.

\subsection{Analysis of ARLM}
\label{ssec:analy}

\begin{conjecture}
	The starting position of the \emph{global maxima} is always present in the
	$start\_pos$ list.
\end{conjecture}
%

\begin{corollary}
	From the above conjecture, it follows that the ending position of the \emph{global maxima} 
	is also present in the $end\_pos$ list.  
\end{corollary}
\begin{proof}
	This directly follows from the commutative property stated in Section~\ref{sec:def}.
\end{proof}

Finding all the local maxima in the string requires a single pass, which takes
$O(l)$ time for a string of length $l$.  Let the number of local maxima in the
string be $d$.  Finding the maximum valued suffix for each local maxima using
the $symbol\_count$ table, requires another pass of each of the local maxima,
and thus also takes $O(l)$ time.  Since, each local maximum contributes one
position to the lists, the number of elements in both the lists is $d$.  In the
rare case that a local maxima contains two or more suffices with the same
maximum $\chi^2$ value greater than that of the local maxima, we store all such
positions in the corresponding list.  Thus, the lists are of $O(d)$ length.  We
then evaluate the substrings formed by each possible pair of start and end
positions, which takes $O(d^2)$.  So in total, the time complexity of the
algorithm becomes $O(l+d^2)$.  

We justified that although $d$ is of $O(l)$, the expected number of local maxima
is far less than that (supported by empirical values shown in
Section~\ref{sec:expt}).  So although the theoretical running time degenerates
to $O(l^2)$, practically it is found to be much better.  The following
optimization further reduces the running time of the algorithm.  We evaluate the
chi-square values only when the substrings are properly formed from the two
lists, i.e., for a given pair of start and end positions obtained from the two
lists, the ending position is greater than or equal to the starting position.
This further reduces the actual running time required compared to that given by
$O(d^2)$.  We empirically show that the running time is actually 3-4 times less
than the na\"\i ve algorithm which computes and compares the value of all the
possible substrings present in the original string.

\subsection{Approximate Greedy Maximum Maxima Search Algorithm (AGMM)}
\label{ssec:greedy}

In this section, we propose a linear time greedy algorithm for finding the 
maximum substring, which is linear in the size of the input string $str$.  
We extract all the local maxima of the input string and its reverse, and 
populate the $start\_pos$ and $end\_pos$ lists as discussed previously.  
We identify the local maxima suffix $max$ having the 
maximum chi-square value among all the local maxima present in the string. AGMM 
assumes this local maxima suffix to be completely present within the global maxima. We 
then find a position $g \in start\_pos$ for which the new substring starting 
at $g$ and ending with $max$ as a suffix has the maximum $\chi^2$ value, for all $g$.  
Using this reconstructed substring, we find a position $h \in end\_pos$ such that 
the new string starting at the selected position $g$ and ending at position $h$ has 
the maximum chi-square measure for all positions of $h$.  This new substring is reported 
by the algorithm as the global maxima.  

\begin{algorithm}[t]
\caption{AGMM Algorithm}
\label{alg:agmm}
\begin{algorithmic}[1]
\REQUIRE String $S$ with $start\_pos$ and $end\_pos$ lists.
\ENSURE Top-k statistically most significant substrings.
\STATE $max \leftarrow$ suffix having the maximum $\chi^2$ value.
\STATE $G \leftarrow$ strings starting at positions from $start\_pos$ with $max$ as suffix.
\STATE $H \leftarrow$ strings in $G$ ending at positions from $end\_pos$.
\STATE Return the top-k strings of $H$ based on $\chi^2$ value.
\end{algorithmic}
\end{algorithm}

Again, using the example of $str=aaaabbba$, we find $max=aaaa$ and using the two lists, 
$aaaa$ is returned as the global maxima.  For $k=2$, the heuristic returns $aaaabbba$ as 
the second most significant substring.

Using the $symbol\_count$ table, AGMM takes $O(d)$ time, where $d$ is the number of local maxima 
found.  The total running time of the algorithm is $O(d+l)$.  However, the substring returned may not 
be the actual global maxima at all times.  The intuition is that the global maxima will 
contain the maximum of the local maxima to maximize its value.  This assumption is justified 
by empirical results in Section~\ref{sec:expt}, which shows that almost always we obtain 
an approximation ratio of 0.96 or more, if not the exact value.  Being a linear time 
algorithm, it provides a order of increase in the runtime as compared to the other algorithms.  
While finding the values of $g$ and $h$, we can keep a track of the chi-squared values of 
all the strings thus formed.  Using these values, the heuristics can be used to output the 
top-k substrings (see Algorithm~\ref{alg:agmm} for the pseudo-code).

\section{Experiments}
\label{sec:expt}

To assess the performance of the two proposed heuristics ARLM and AGMM, we conduct tests on
multiple datasets and compare it with the results of the na\"\i ve algorithm and the
blocking algorithm~\cite{sumit}.  The heap variant of the blocking algorithm is
not efficient as it has a higher running time complexity and uses more memory,
and hence has not been compared with.  The accuracy of the results returned by
the heuristics is compared with that returned by the na\"\i ve algorithm, which
provides the optimal answer.  

We used two real datasets: (i)~innings by innings runs scored by Sachin
Tendulkar in one-day internationals (ODI)\footnote{
\url{http://stats.cricinfo.com/ci/engine/player/35320.html?}\\\url{class=2;template=results;type=batting;view=innings}.}
and (ii)~the number of user clicks on the front page of
\url{msnbc.com}\footnote{
\url{http://archive.ics.uci.edu/ml/}\\\url{datasets/MSNBC.com+Anonymous+Web+Data}.}.
We have also used synthetic data to assess the scalability and practicality of
the heuristics.  We compare the results based on the following parameters:
(i)~search time for top-k queries, (ii)~number of local maxima found, and
(iii)~accuracy of the result based on the ratio of the optimal $\chi^2$ value
obtained from the na\"\i ve algorithm to that returned by the algorithms.  The
experiments were conducted on a 2.1 GHz desktop PC with 2 GB of memory using C++
in Linux environment.

\subsection{Real Datasets}

\begin{table}[t]
	\begin{center}
		\begin{small}
		\begin{tabular}{|c|c|c|c|c|c|}
			\hline
			{\bfseries \# innings} & {\bfseries Total runs} & {\bfseries Avg.} & 
			{\bfseries \# 100} & {\bfseries \# 50} & {\bfseries \# 0} \\
			\hline
			\hline
			425 & 17178 & 44.50 & 45 & 91 & 20 \\
			\hline
		\end{tabular}
		\caption{Sachin Tendulkar's ODI career statistics (as on November, 2009).}
		\label{tab:sachin}
		\end{small}
	\end{center}
\end{table}

Table~\ref{tab:sachin} summarizes the statistics of Sachin Tendulkar's present
ODI career.  The innings where he did not bat were not considered.  Given his
runs, we quantized the runs scored into 5 symbols as follows: 0-9 is represented
by $A$ (Poor), 10-24 by $B$ (Bad), 25-49 by $C$ (Fair), 50-99 by $D$ (Good) and
100+ by $E$ (Excellent).  His innings-wise runs were categorized, and from the
entire data we calculated the actual probability of occurrences of the different
symbols, which were $0.28$, $0.18$, $0.22$, $0.22$ and $0.10$ respectively for
the five symbols.  With this setting, we extracted the top-k substring with the
maximum chi-square value.  These results reflect the periods of his career when
he was in top form or when there was a bad patch, since in both cases his
performance would deviate from the expected.  Table~\ref{tab:sachinres}
summarizes the findings.  We find that during his best patch he had scored $8$
centuries and $3$ half-centuries in 20 innings with an average of $84.31$, while
in the off-form period he had an average of $21.89$ in 9 innings without a score
above 40.  

\begin{table}[t]
	\begin{center}
		\begin{small}
		\begin{tabular}{|c|c|c|c|}
			\hline
			{\bfseries Form} & {\bfseries Date} & {\bfseries Avg.} & {\bfseries Runs scored} \\
			\hline
			\hline
			& {22/04/1998} & & {143,134,33,18,100*} \\
			{\bfseries Best} & {to} & {84.31} & {65,53,17,128,77} \\
			{\bfseries patch }& {13/11/1998} & & {127*,29,2,141,8,3} \\
			& & & {118*,18,11,124*} \\
			\hline
			{\bfseries Worst} & {15/03/1992} & & {14,39,15} \\
			{\bfseries patch} & {to} & {21.89} & {10,22,21} \\
			& {19/12/1992} & & {32,23,21}\\
			\hline
		\end{tabular}
		\caption{Result from Sachin's records.}
		\label{tab:sachinres}
		\end{small}
	\end{center}
\end{table}

\begin{figure}[t]
	\begin{center}
		\includegraphics[width=\figwidth]{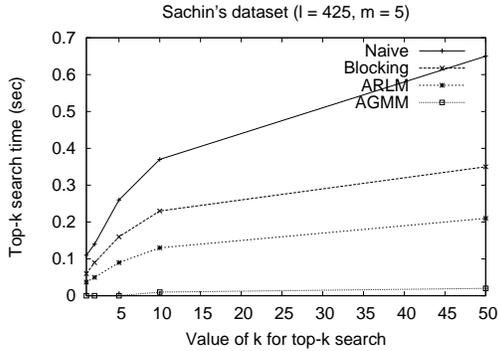} 
		\caption{Time for finding the top-k query in Sachin's run dataset.}
		\label{grap:sachin_time}
	\end{center}
\end{figure}

\begin{figure}[t]
	\begin{center}
		\includegraphics[width=\figwidth]{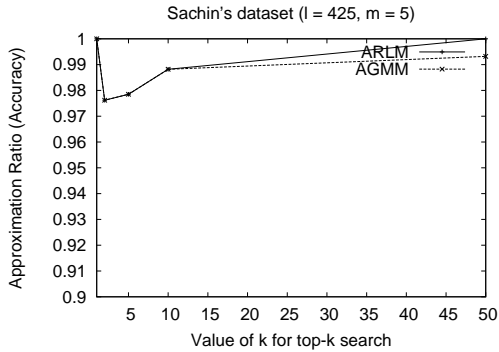}
		\caption{Approximation ratio of the top-k query in Sachin's run dataset.}
		\label{grap:sachin_timeb}
	\end{center}
\end{figure}

Figure~\ref{grap:sachin_time} and Figure~\ref{grap:sachin_timeb} plot the times
taken by the different algorithms and the approximation factor or accuracy of
result for the heuristics respectively while varying the values of top-k
queries.  The ARLM algorithm takes lesser running time as compared to the other
procedures, while the AGMM method, being a linear time algorithm, is very fast.
The accuracy of the ARLM heuristic is found to be 100\% for the top-1 query,
i.e., it provides the correct result validating the conjecture we proposed in
Section~\ref{sec:algo}.  As the value of $k$ increases we find an increase in
the approximation ratio of both the heuristics as the number of pairs of local
maxima involved increases, giving better results.  The number of local maxima
found is lesser than the number of blocks constructed by the blocking algorithm
(see Table~\ref{tab:comp}).  So, the heuristic prunes the search space more
efficiently.  

The second real dataset that we considered contained the number of user clicks
encountered on the front page of the website \url{msnbc.com} during various
periods of a day taken from a sample of 989819 users.  Analysis of the clicks
from a group of users provides an insight into potent clients for the
organization or customers for e-commerce purposes.  The number of clicks have
been categorized as follows: 1-3 clicks have been represented by $A$ (Low), 4-9
clicks by $B$ (Medium) and 10+ clicks by $C$ (High).  We accordingly quantized
the dataset and then performed the experiments by calculating the actual
probability of occurrences of the different symbols which were $0.43$, $0.36$
and $0.21$ respectively.  Table~\ref{tab:click} describes the data values and
tabulates the result for the top-1 query.  Due to time-consuming nature of this
dataset, we did not search for the top-k queries with varying values of $k$.
The results show that the ARLM technique has a better running time than the
others, and also operates on a lesser number of local maxima as opposed to the
number of blocks for the blocking algorithm (see Table~\ref{tab:comp}).  The
approximation factor for both the heuristics is $1$ for the top-1 search,
thereby yielding the correct result. 

\begin{table}[t]
	\begin{center}
		\begin{small}
		\begin{tabular}{|c|c|c|}
			\hline
			{\bfseries Algorithm} & {\bfseries Searching time} & {\bfseries	Approx. ratio} \\
			\hline
			\hline
			{Na\"\i ve} & {75+ hrs} & 1 \\
			\hline
			{Blocking} & {52 hrs} & 1 \\
			\hline
			{ARLM} & {40 hrs} & 1 \\
			\hline
			{AGMM} & {3 hrs} & 1 \\
			\hline
		\end{tabular}
		\caption{Results for dataset (containing 989819 records) of number of user clicks.}
		\label{tab:click}
		\end{small}
	\end{center}
\end{table}

\begin{table}[t]
	\begin{center}
		\begin{small}
		\begin{tabular}{|c|c|c|}
			\hline
			{\bfseries Dataset} & {\bfseries \# Blocks} & {\bfseries \# Local maxima} \\
			\hline
			\hline
			{Sachin} & {319} & {281} \\
			\hline
			{Web clicks} & {835142} & {759921} \\
			\hline
		\end{tabular}
		\caption{Number of blocks versus local maxima for real datasets.}
		\label{tab:comp}
		\end{small}
	\end{center}
\end{table}

\subsection{Synthetic datasets}

We now benchmark the ARLM and AGMM heuristics against datasets generated
randomly using a uniform distribution.  To simulate the deviations from the expected
characteristics as observed in real applications, we perturb the random data thus
generated with chunks of data generated from a geometric distribution with parameter $p=0.3$.  These strings are
now mined to extract the top-k substrings with largest chi-square values.  The
parameters that affect the performance of the heuristics are: (i)~length of the
input string, $l$, (ii)~size of the alphabet set, $m$, and (iii)~number of top-k values.  
For different values of these parameters we compare our
algorithms with the existing ones on the basis of (a)~time to search,
(b)~approximation ratio of the results, and (c)~the number of blocks evaluated in
case of blocking algorithm to the number of local maxima found by our
algorithm.

\subsection{Effect of parameters}

\begin{table}[t]
	\begin{center}
		\begin{small}
		\begin{tabular}{|c|c|c|c|}
			\hline
			{\bfseries Parameters} &{\bfseries Variable} & {\bfseries \# Blocks} & {\bfseries \# Local maxima} \\
			\hline
			\hline
			{m=5,} & {l=$10^3$} & {831} & {742} \\
			{k=1} & {l=$10^4$} & {7821} & {6740} \\
			& {l=$10^5$} & {77869} & {66771} \\
			\hline
			{l=$10^4$,} & {m=5} & {7821} & {6740} \\
			{k=1} & {m=25} & {8104} & {7203} \\
			& {m=50} & {8704} & {7993} \\
			\hline
		\end{tabular}
		\caption{Results for uniform dataset.}
		\label{tab:unif}
		\end{small}
	\end{center}
\end{table}

\begin{figure}[t]
	\begin{center}
		\includegraphics[width=\figwidth]{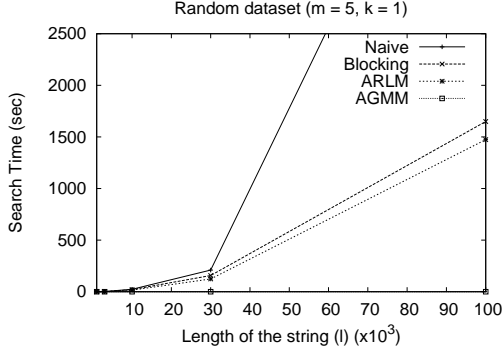}
		\caption{Effect of length on search time.}
		\label{grap:time_l}
	\end{center}
\end{figure}

\begin{figure}[t]
	\begin{center}
		\includegraphics[width=\figwidth]{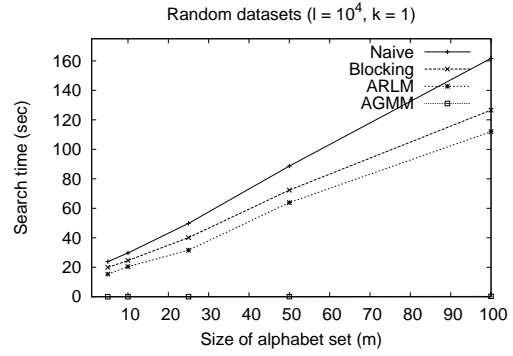}
		\caption{Effect of size of alphabet size on search time.}
		\label{grap:time_lb}
	\end{center}
\end{figure}

\begin{figure}[t]
	\begin{center}
		\includegraphics[width=\figwidth]{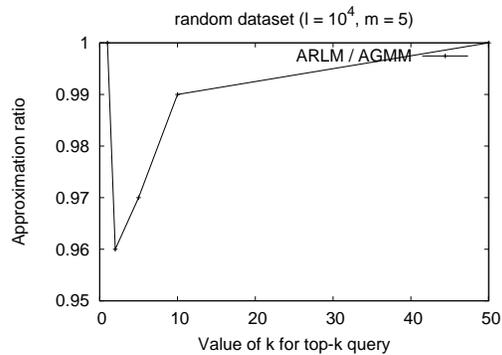}
		\caption{Approximation ratio of the top-k query.}
		\label{grap:acc_k}
	\end{center}
\end{figure}

Figure~\ref{grap:time_l} shows that with the increase in the length of the input
string $l$, the time taken for searching the top-k queries increases.  The
number of blocks or local maxima increases with the size of the string and hence
the time to compute also increases.  The time increases more or less
quadratically for ARLM and the other existing algorithms according to the
analysis shown in Section~\ref{ssec:analy}.  ARLM takes less running time than
the other techniques, as the number of local maxima found is less than the
number of blocks found by the blocking algorithm (see Table~\ref{tab:unif}).
Hence, it provides better pruning of the search space and is faster.  On the
other hand, AGMM being a linear time heuristic runs an order of time faster than
the others.  We also find that the accuracy of the top-k results reported by
AGMM show an improvement with the increase in the string length (see
Figure~\ref{grap:acc_k}), as the deviation of substrings become more prominent
with respect to the large portions of the string depicting expected behavior.
The approximation factor for ARLM is $1$ for the top-$1$ query in all the cases
tested, while for other top-k queries and for AGMM it is always above $0.96$. \\

\begin{figure}[t]
	\begin{center}
		\includegraphics[width=\figwidth]{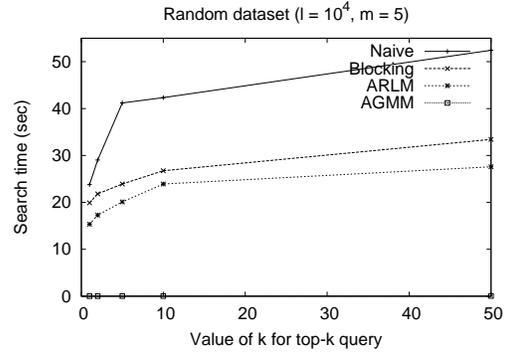}
		\caption{Effect of value of k for top-k query on search time.}
		\label{grap:time_pb}
	\end{center}
\end{figure}

\begin{figure}[t]
	\begin{center}
		\includegraphics[width=\figwidth]{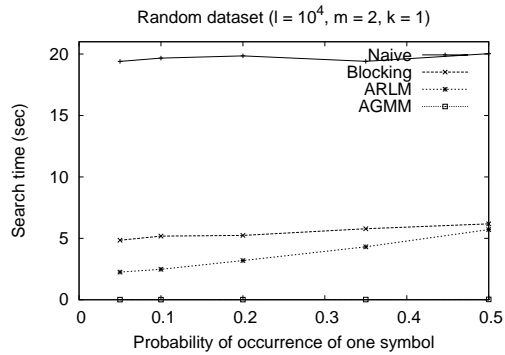}
		\caption{Effect of probability in two symbol string on search time.}
		\label{grap:time_p}
	\end{center}
\end{figure}

Varying the size of the alphabet set $m$, we find that the time taken for
searching the top-k query as well as the number of blocks formed increases
(Table~\ref{tab:unif} and Figure~\ref{grap:time_lb}).  As $m$ increases, the
number of blocks increases as the probability of the same symbol occurring
contiguously falls off.  We have observed in Section~\ref{sec:def} that a local
maxima can only end at positions containing adjacent dissimilar symbols.  So the
number of local maxima found increases, thereby increasing the computation time
of the algorithms.  There seems to be no appreciable effect of $m$ on the
approximation ratio of the results returned by the algorithms.  We tested with
varying values of $m$ with $l=10^4$ and $k=2$, and found the ratio to be $1$ in
all cases.  Figure~\ref{grap:time_p} shows the effect of varying probability of
occurrence of one of the symbols in a string composed of two symbols only.  The
approximation ratio remained $1$ for both heuristics for the top-$1$ query. \\

We next show the scalability of our algorithms by conducting experiments for
varying values of $k$ for top-k substrings.  Figure~\ref{grap:time_pb} shows that
search time increases with the increase in the value of $k$.  This is evident as
we are required to perform more computations.  The accuracy of the results for
the heuristics increases with $k$.  For $k = 2$, it is 0.96, and increases up to
1 when $k$ becomes more than 10.  The number of blocks or local maxima found
remains unchanged with the variation of $k$.

\section{Conclusions}
\label{sec:conc}

In this paper, we have proposed two heuristics for searching a given string for
the top-k substrings having the maximum chi-square value representing its
deviation from the expected nature, with the possibility of hidden pattern or
information.  We described how the chi-square measure closely approximates
p-value and is apt for mining such substrings.  We provided a set of
observations based on which we developed two heuristics, one which runs in time
quadratic with the number of local maxima, and the other which is linear.  Our 
experiments showed that the proposed heuristics are faster
than the existing algorithms.  The algorithms return results that have an
approximation ratio of more than 0.96.

{
\bibliographystyle{abbrv}
\bibliography{ref}

\begin{thebibliography}{10}

\bibitem{sumit}
S.~Agarwal.
\newblock On finding the most statistically significant substring using the
  chi-square measure.
\newblock Master's thesis, Indian Institute of Technology, Kanpur, 2009.

\bibitem{mot}
D.~Bainchi and B.~Tirozzi.
\newblock Identifying short motifs by means of extreme value analysis.
\newblock {\em Europhysics Letters}, 84(1):18001p1--18001p6, 2008.

\bibitem{categ}
G.~Bejerano, N.~Friedman, and N.~Tishby.
\newblock Efficient exact p-value computation for small sample, sparse and
  surprisingly categorical data.
\newblock {\em J. Computational Biology}, 11(5):867--886, 2004.

\bibitem{bioinfo}
A.~Denise, M.~Regnier, and M.~Vandenbogaert.
\newblock Accessing the statistical significance of overrepresented
  oligonucleotides.
\newblock In {\em Workshop on Algorithms in Bioinformatics (WABI)}, pages
  85--97, 2001.

\bibitem{hotel}
H.~Hotelling.
\newblock Multivariate quality control.
\newblock {\em Techniques of Statistical Analysis}, 54:111--184, 1947.

\bibitem{keo}
E.~Keogh, S.~Lonardi, and B.~Chiu.
\newblock Finding surprising patterns in a time series database in linear time
  and space.
\newblock In {\em Proc. of 8th ACM SIGKDD Int. Conf. on Knowledge Discovery and
  Data Mining}, pages 550--556, 2002.

\bibitem{dp}
S.~Rahmann.
\newblock Dynamic programming algorithms for two statistical problems in
  computational biology.
\newblock In {\em Workshop on Algorithms in Bioinformatics (WABI)}, volume 2812
  of {\em LNCS}, pages 151--164, 2003.

\bibitem{multi}
T.~Read and N.~Cressie.
\newblock {\em Goodness-of-fit statistics for discrete multivariate data}.
\newblock Springer, 1988.

\bibitem{pear}
T.~Read and N.~Cressie.
\newblock Pearson's {$\chi^2$} and the likelihood ratio statistic {$G^2$}: a
  comparative review.
\newblock {\em International Statistical Review}, 57(1):19--43, 1989.

\bibitem{crit}
M.~Regnier and M.~Vandenbogaert.
\newblock Comparison of statistical significance criteria.
\newblock {\em J. Bioinformatics and Computational Biology}, 4(2):537--551,
  2006.

\bibitem{anomaly}
N.~Ye and Q.~Chen.
\newblock An anomaly detection technique based on chi-square statistics for
  detecting intrusions into information systems.
\newblock {\em Quality and Reliability Engineering International},
  17(2):105--112, 2001.

\end{thebibliography}
}

\end{document}